\documentclass[aps,prl,twocolumn,superscriptaddress,longbibliography]{revtex4-2}
\usepackage{graphicx}
\usepackage{latexsym}
\usepackage{amssymb}
\usepackage{amsmath}
\usepackage{amsfonts}
\usepackage{upgreek}
\usepackage{float}
\usepackage{bm}
\usepackage{multirow}
\usepackage{color}
\usepackage[T1]{fontenc}
\usepackage{hyperref}
\usepackage{subfigure}
\usepackage{booktabs,tabularx}

\hypersetup{
colorlinks = true,
linkcolor = [rgb]{0.70,0.13,0.13},
citecolor = [rgb]{0.13,0.55,0.13},
urlcolor  = [rgb]{0.25, 0.41, 0.88}}

\begin{document}


\title{Fidelity zeros and Lee-Yang theory of quantum phase transitions}

\author{Tian-Yi Gu}
\affiliation{College of Physics, Nanjing University of Aeronautics and Astronautics, Nanjing, 211106, China}
\affiliation{Key Laboratory of Aerospace Information Materials and Physics (NUAA), MIIT, Nanjing 211106, China}

\author{Gaoyong Sun}
\thanks{Corresponding author: gysun@nuaa.edu.cn}
\affiliation{College of Physics, Nanjing University of Aeronautics and Astronautics, Nanjing, 211106, China}
\affiliation{Key Laboratory of Aerospace Information Materials and Physics (NUAA), MIIT, Nanjing 211106, China}

\begin{abstract}
Lee-Yang theory is central to the analysis of thermal phase transitions. However, the underlying mechanism of the theory and the nature of Lee-Yang zeros in quantum many-body systems remains elusive. Here, we develop a unified framework for understanding quantum phase transitions from fidelity zeros induced by symmetry breaking. These zeros, arising from transitions between symmetry sectors, obey the Lee-Yang theorem and give rise to fidelity edges near critical points. Quantum criticality is further characterized through the finite-size scaling of fidelity zeros. As concrete examples, we investigate fidelity zeros analytically in one-dimensional and numerically in two-dimensional quantum Ising models under a complex magnetic field, showing excellent agreement with analytical theory and with quantum Monte Carlo and tensor-network simulations. Our results provide new insights into the mechanism of Lee-Yang theory and open avenues for exploring unexplored landscapes of phase transitions in quantum many-body systems.

\end{abstract}

\maketitle

{\bf Introduction.}
The study of phase transitions lies at the heart of statistical and condensed matter physics, revealing the nature of many-body systems \cite{sachdev1999quantum}. 
Thermal (classical) phase transitions are typically characterized by zeros of the partition function and singularities in the free energy in the thermodynamic limit.
However, for a finite system, the partition function is always a positive and analytic function for real values of the control parameter.
To reveal the underlying mechanism of phase transitions, Yang and Lee developed the seminal framework now known as Lee-Yang theory, which extends the control parameter into the complex plane and analyzes the zeros (dubbed Lee-Yang zeros) of the partition function in finite classical Ising models \cite{yang1952statistical,lee1952statistical}. 
In the canonical ensemble, Fisher demonstrated that phase transitions can be characterized by the zeros of the partition function in the complex inverse-temperature plane \cite{fisher1965statistical}, leading to the concept of Fisher zeros. 
The Lee-Yang framework has likewise been extended to dynamical quantum phase transitions \cite{heyl2013dynamical,heyl2018dynamical}, yielding the notion of Loschmidt zeros.
Today, Lee-Yang theory stands as a cornerstone of our understanding of phase transitions, not only giving rise to the celebrated Lee-Yang theorem \cite{lee1952statistical} but also inspiring modern approaches to critical phenomena, such as Yang-Lee edge singularity, based on nonunitary conformal field theory \cite{fisher1978yang,kortman1971density,cardy1985conformal}.

\begin{figure}[t]
\includegraphics[width=8.6cm]{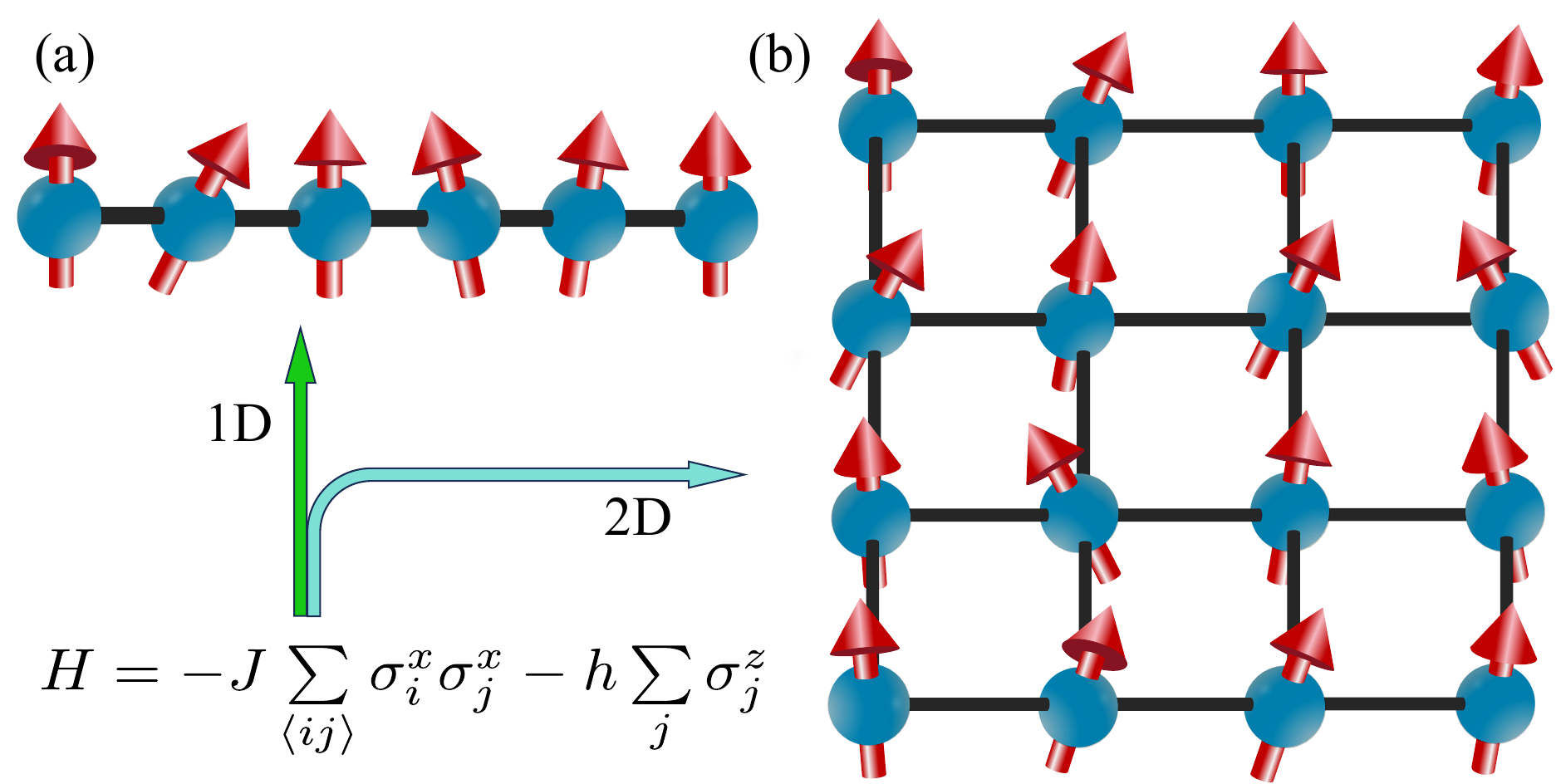} 
\centering
\caption{ Geometric structures of one dimensional (1D) and two dimensional (2D) transverse-field quantum Ising models. (a) Quantum Ising model in a chain, (b) Quantum Ising model in a square lattice. }
\label{fig:model}
\end{figure}

Lee-Yang zeros in the complex plane of control parameters were initially regarded as unphysical \cite{ananikian2015imaginary,peng2015experimental}, hereby limiting early studies to indirect observations \cite{binek1998density,binek2001yang}. 
Their first direct experimental observation \cite{peng2015experimental} was realized through the spin coherence measurement \cite{wei2012lee}.
Recently, rapid progress in non-Hermitian physics  \cite{bergholtz2021exceptional,ashida2020non,yao2018edge,yokomizo2019non,yang2020non,okuma2020topological,zhang2020correspondence,lin2023topological} have enabled experimental observations \cite{gao2024experimental,lan2024experimental} of Yang-Lee edge singularity in open systems. 
Together with the experiments of dynamical Lee-Yang zeros \cite{brandner2017experimental} and dynamical quantum phase transitions \cite{jurcevic2017direct,flaschner2018observation,xu2020measuring,wang2019simulating,tian2019observation,guo2019observation,nie2020experimental,tian2020observation,wu2024indication}, these advances have significantly stimulated the development of the field.
A fundamental question is how to probe quantum phase transitions within the Lee-Yang framework in correlated many-body systems.
By exploring the partition function or introducing moment-generating function, the Lee-Yang formalism has been extended to characterize quantum phase transitions \cite{kist2021lee,vecsei2022lee,vecsei2023lee,liu2024imaginary,liu2024exact,vecsei2025lee,li2023yang,li2025yang}, many-body dynamics \cite{meng2025detecting,wang2024quantum} and the fermion sign problem \cite{he2025revisiting}.
However, so far, the underlying mechanism of Lee-Yang theory and the nature of Lee-Yang zeros in quantum many-body systems remains elusive.

We present a unified Lee-Yang framework based on fidelity zeros, which serve as Lee-Yang/Fisher zeros in classical phase transitions and as Loschmidt zeros in dynamical quantum phase transitions. We show that fidelity zeros emerge exclusively in the ordered phase and disappear abruptly at the phase transition, vanishing in the disordered phase. These zeros arise from symmetry breaking triggered by a complex magnetic field and are proven to satisfy the Lee-Yang theorem. They form branch points, termed fidelity edges that are directly analogous to the Yang-Lee edge singularity. As the control parameter approaches the critical point, the fidelity edges pinch the real axis, signaling quantum phase transitions. We further develop a finite-size scaling theory near the critical point, with numerical results in excellent agreement with quantum Monte Carlo and tensor-network simulations. Our findings provide new insights into the Lee-Yang theory and establish fidelity zeros as a powerful numerical tool for probing quantum phase transitions in correlated many-body systems.

\begin{figure}[t]
\includegraphics[width=8.3cm]{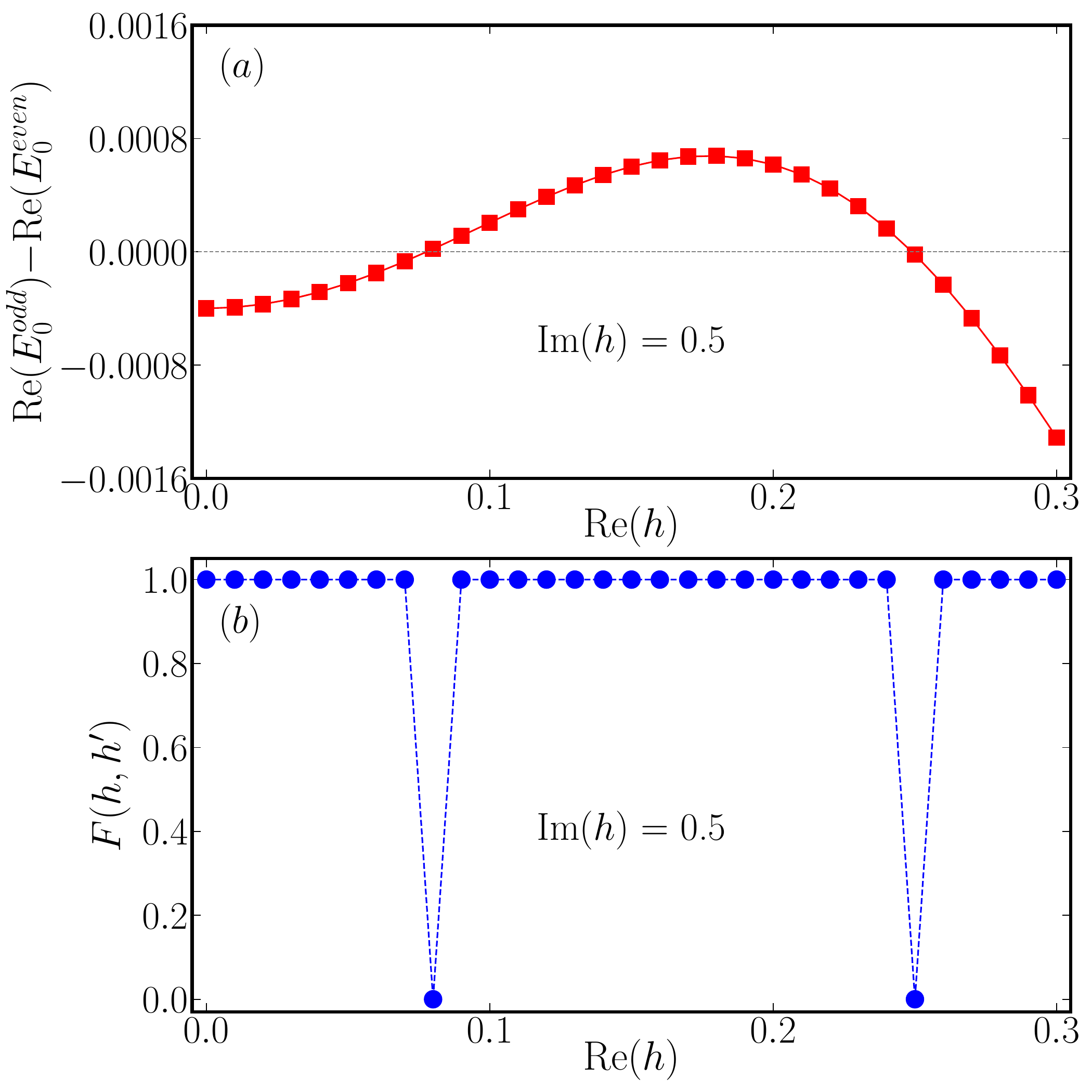} 
\centering
\caption{Non-Hermitian parity-symmetry breaking. (a) Real-part energy difference $\text{Re}(E_{0}^{\text{odd}})-\text{Re}(E_{0}^{\text{even}})$ between the odd- and even-parity symmetric ground state as the function of $\text{Re}(h)$ for $L=10$ at $\text{Im}(h) =0.5$, (b) Fidelity $F(h,h^{\prime})$ with $\delta h=0.01$ as a function of $\text{Re}(h)$ under the same parameters as in (a). Here, the solid symbols represent the numerical results, and the solid and dashed lines act as eye guides. }
\label{fig:paritybreak}
\end{figure}

{\bf Non-Hermitian symmetry breaking.} 
To illustrate our framework, we first consider the one-dimensional (1D) quantum Ising model, 
\begin{equation}
H = - J \sum_{\langle i, j \rangle} \sigma_{i}^{x} \sigma_{j}^{x} - h \sum_{j} \sigma_{j}^{z},
\label{Eq.Ham}
\end{equation}
as shown in Fig.~\ref{fig:model}(a), subjected to a complex transverse field along the $z$-direction.
Here, $\sigma_{j}^{x}$ and $\sigma_{j}^{z}$ are Pauli matrices along $x$ and $z$ directions at the $j$th lattice site. $J>0$ is a real coupling constant, and $h$ denotes a complex magnetic field. 
The notation \(\langle i, j \rangle\) denotes a summation over all pairs of neighboring sites. 
Throughout the paper, we impose $J=1$ and periodic boundary conditions, $\sigma_{L+1}^x = \sigma_{1}^x$ with $L$ is the length of the chain.
For a purely real magnetic field $\text{Im}(h)$=0, the system is the conventional Hermitian quantum Ising chain, which undergoes a quantum phase transition at $\text{Re}(h) = 1$ from the ferromagnetic phase to the paramagnetic phase. For $\text{Im}(h) \neq 0$, the Hamiltonian is yet non-Hermitian.

Although the particle number (equivalently, the total $z$-magnetization in spin language) is not conserved, the Hamiltonian (\ref{Eq.Ham}) is invariant under the parity symmetry ($\mathbb{Z}_2$ symmetry) generated by $P = \prod_{j=1}^{L} \sigma_{j}^{z}$. Accordingly, the Hilbert space decomposes into odd- and even-parity sectors, projected by $P_{\text{odd}} = \tfrac{1}{2}(1-P)$ and $P_{\text{even}} = \tfrac{1}{2}(1+P)$, respectively \cite{mbeng2024quantum}. Consequently, the ferromagnetic phase is doubly degenerate in the thermodynamic limit.
Through the Jordan-Wigner and Bogoliubov transformations, Hamiltonian (\ref{Eq.Ham}) is diagonalized, yielding the eigenvalues,
$\epsilon(k) = 2J\sqrt{(\cos k - \frac{h}{J})^{2} + \sin^{2}k}$,
where $k=2n\pi/L$ with $n=-L/2+1, \cdots, L/2$ for odd-parity states, and $k=\pm(2n-1)\pi/L$ with $n=1, \cdots, L/2$ for even-parity states.
For a real field \text{Im}(h)=0, the odd- and even-parity ground-state energies are obtained by occupying single-particle energy levels. 
Despite double degeneracy of the ferromagnetic phase in the thermodynamic limit, these states are separated by an exponentially small gap for finite $L$.
It is observed that the odd-parity ground state always lies at a lower energy.
Notably, the Hamiltonian's parity symmetry remains preserved despite the presence of non-Hermiticity. 
However, we find that the ground state (defined as the lowest real part of energies \cite{sun2023aufbau}) is not always of odd parity but can switch symmetry sector [c.f. Fig.~\ref{fig:paritybreak}(a)] under a complex magnetic field $\text{Im}(h) \neq 0$. We refer to this type of parity breaking as {\it non-Hermitian parity-symmetry breaking}.

In classical Ising models, the partition function is nonzero for any real magnetic field in a finite system.
In Lee-Yang’s original framework, a complex magnetic field is introduced to generate the zeros of the partition function.
Thermal phase transitions are identified when these zeros cross the real axis in the thermodynamic limit.
In what follows, we demonstrate that non-Hermitian parity-symmetry breaking caused by the complex field in quantum Ising models leads to fidelity zeros, thereby establishing a generalized Lee-Yang theory for quantum phase transitions.

\begin{figure}[t]
\includegraphics[width=8.7cm]{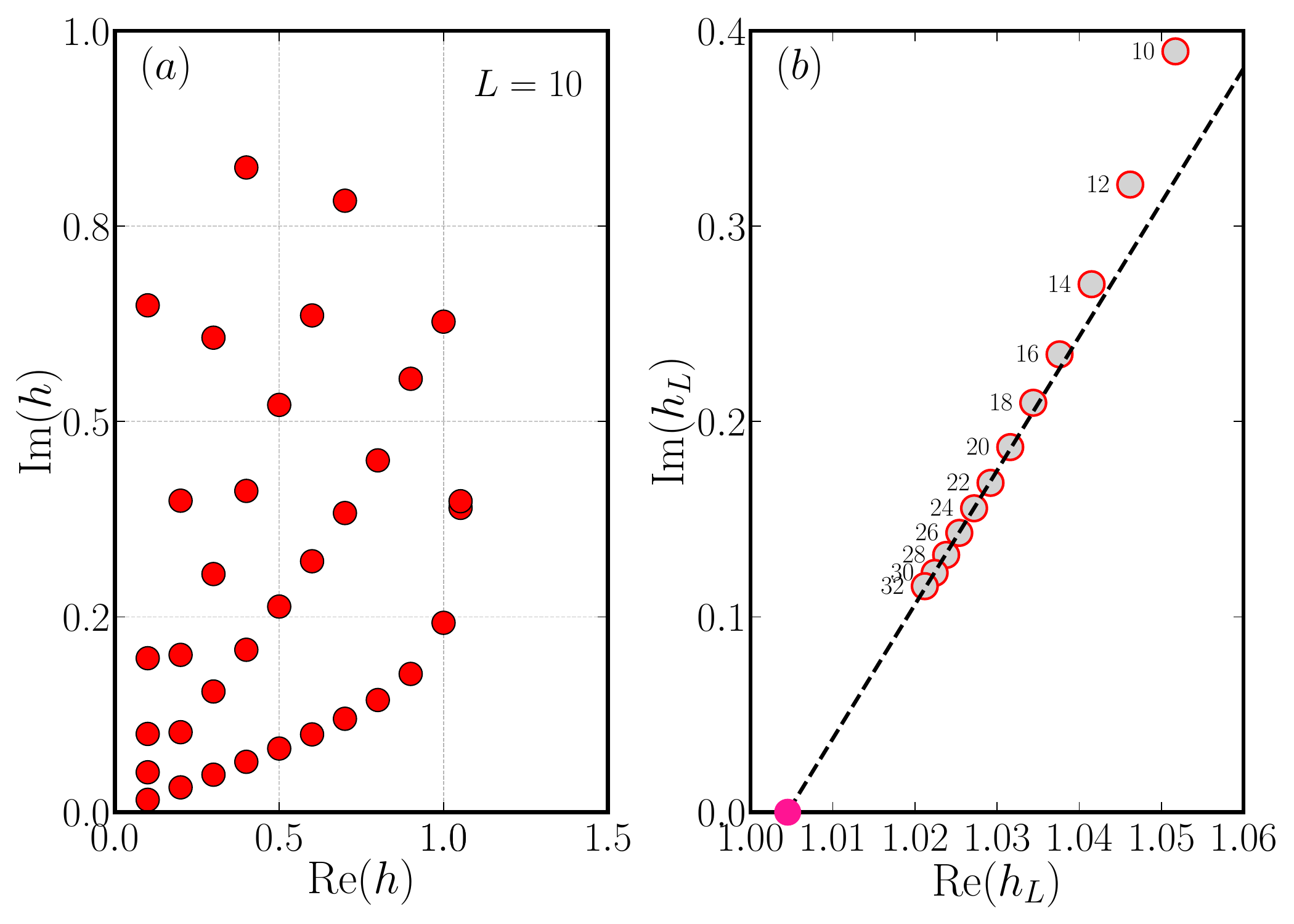} 
\centering
\caption{Fidelity zeros and finite-size scaling of 1D transverse-field quantum Ising model. (a) Distribution of fidelity zeros in the complex plane $h$ with $L=10$ and $\delta h=10^{-4}$. The scatter red points mark the complex field values at which the fidelity vanishes. (b) Finite-size scaling of the fidelity-zero positions $h_{L}$ as a function of system size $N$ ranging from $10$ to $32$. Here, The grey-red dots denote the complex field values possessing the largest real and smallest imaginary parts. The black dashed line represents the fitted curve. The pink solid dot on the real axis marks the critical value $h_{c1} = 1.0045$, consistent with the theoretical prediction. }
\label{Fig:1Dcritical}
\end{figure}

{\bf Fidelity zeros and scaling laws.} 
To characterize changes between different parity-symmetry sectors, we introduce the ground-state fidelity \cite{gu2010fidelity,you2007fidelity}, $F(h, h^{\prime}) = |\langle \Psi_0(h') | \Psi_0(h) \rangle|$, 
which quantifies the overlap of the ground states $|\Psi_0(h)\rangle$ and $|\Psi_0(h^{\prime})\rangle$ at fields $h$ and $h^{\prime}=h+\delta h$. 
For Hamiltonian (\ref{Eq.Ham}), the ground-state fidelity is given exactly by,
\begin{equation}
F(h, h^{\prime}) = \prod_{k>0} |( \cos^{\ast}\theta_{k}^{\prime} \cos\theta_k + \sin^{\ast}\theta_{k}^{\prime} \sin\theta_{k} )|,
\end{equation}
with the Bogoliubov angle $\theta_{k}$ satisfying the condition $\tan2\theta_k = \sin k/(\cos k - h)$ \cite{gu2010fidelity}. 
We analyze the ground-state fidelity of the model for $L = 10$ and $\text{Im}(h)=0.5$, and present the results in Fig.~\ref{fig:paritybreak}(b). 
The exact fidelity zeros are observed to coincide precisely with the transitions between symmetry sectors.

Typically, the fidelity does not vanish exactly under small variations of a control parameter in finite Hermitian systems, except in special cases. 
In systems with a spatial-lattice interpretation, the vanishing fidelity of a single momentum mode can lead to zeros in the fidelity \cite{zeng2024exact}. 
However, this approach is difficult to extend to correlated systems \cite{zeng2024exact}, such as the quantum Ising model.
In contrast, the exact zeros we identified lie on the complex plane, which provides a general framework for exploring Lee-Yang theory.
 
To further elucidate the behavior of fidelity zeros, we analytically compute the fidelity of 1D transverse-field Ising model with $L = 10$ and depict the resulting distribution in Fig.~\ref{Fig:1Dcritical}(a). As expected, the fidelity zeros appear exclusively in the ferromagnetic phase, with none observed in the paramagnetic phase. 
This abrupt change in their distribution near $\text{Re}(h_L) \approx 1$ strongly indicates that the quantum phase transition occurs at $\text{Re}(h) = 1$. 
We define the finite-size critical point $h_{L}$ as the complex field $h$ having the largest real part and the smallest imaginary part. 
Within the Lee-Yang framework, as the system size $L$ approaches infinity ($L \to \infty$), the complex $h_{L}$ converges toward the real axis, indicating a phase transition at the real field value.

To determine the critical point in the thermodynamic limit, we compute the fidelity zeros and obtain the corresponding critical values $h_{L}$ for different lattice sizes. 
We then fit both $\text{Re}(h_L)$ and $\text{Im}(h_L)$ using the scaling relation, 
\begin{equation}
h_{L} = h_{\infty} + a L^{-1/\nu}, 
\label{Eq:scaling}
\end{equation}
where $h_{\infty}$ and $a$ are fitting parameters, and $\nu$ is the critical exponent of the correlation length. 
We find that the real part converges to unity and the imaginary part tends to zero [c.f. Fig.~\ref{Fig:1Dcritical}(b)], yielding the critical point $h_{c1} = 1.0045$ with $\nu=1$,  in excellent agreement with theoretical predictions, thereby confirming the accuracy and reliability of our numerical results based on the Lee-Yang formalism.

\begin{figure}[t]
\includegraphics[width=8.6cm]{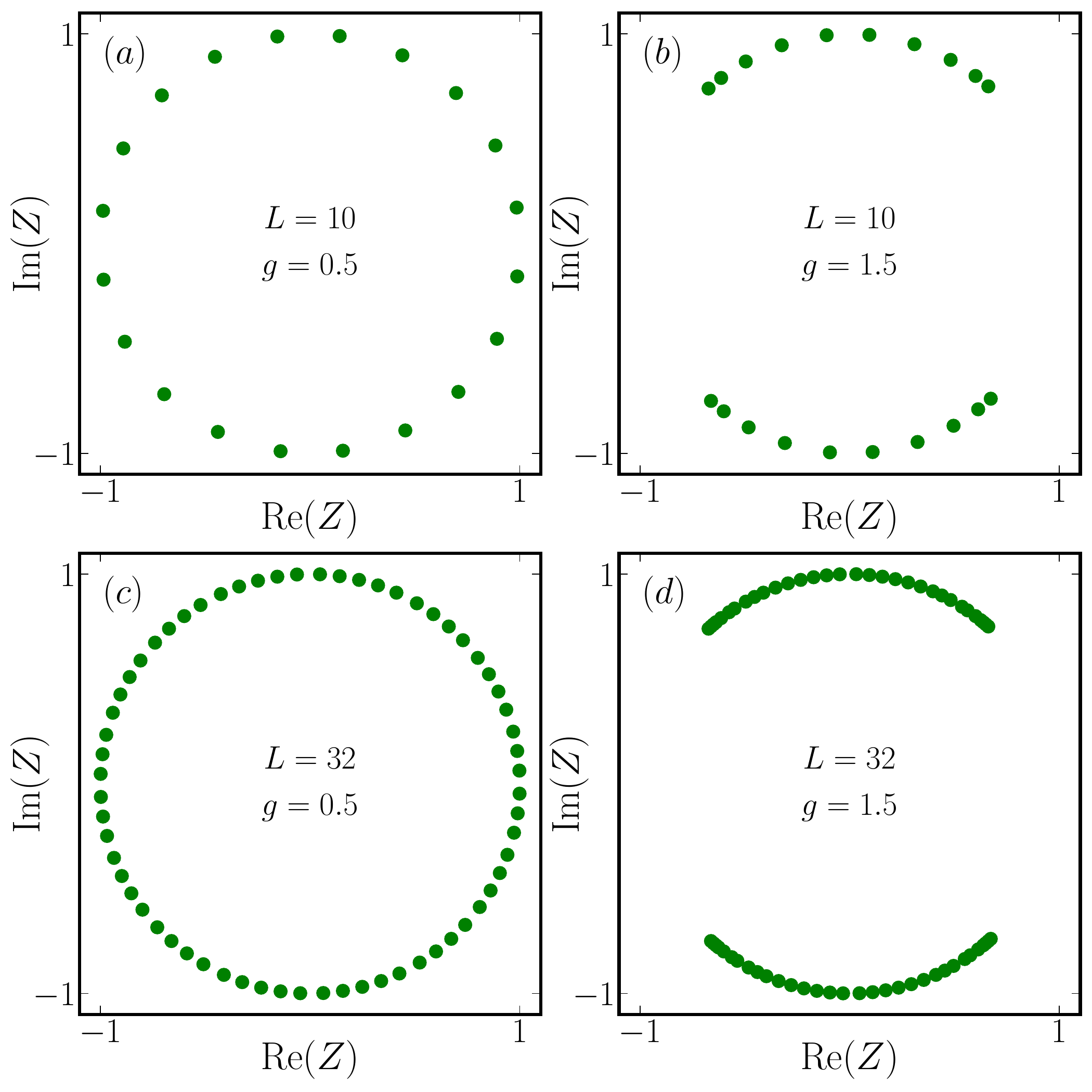} 
\centering
\caption{Fidelity zeros and fidelity edges of 1D quantum Ising model. (a) and (b) Distributions of fidelity zeros in the complex field plane for $L=10$ at $g=0.5$ and $g=1.5$. (c) and (d) Distributions of fidelity zeros in the complex field plane for $L=32$ at $g=0.5$ and $g=1.5$. All distributions of fidelity zeros are located at the unit circle $Z = e^{i\varphi}$ with $\varphi = [0, 2\pi)$. For $g=1.5$, branch points dubbed fidelity edges exist. Here $\delta \varphi = 2\pi/400$ is used for computation of fidelity.}
\label{Fig:YLES1D}
\end{figure}

{\bf Lee-Yang theorem and fidelity edges.}  
In the classical ferromagnetic Ising model, Lee and Yang proved \cite{lee1952statistical} that the zeros of the partition function occur exclusively at purely imaginary values of the external field. In terms of the complex fugacity $z=e^{-2 \beta h}$, these zeros are located exactly on the unit circle, a result now known as the Lee-Yang theorem. 
In the thermodynamic limit, they show that the zeros pinch the real axis below the critical temperature, whereas above it they remain bounded, leaving a gap that prevents them from reaching the real axis and giving rise to the critical phenomenon dubbed the Yang-Lee edge singularity.

To analyze the distribution of fidelity zeros, we introduce the fugacity $Z=e^{i \varphi}$ in the quantum Ising model by expressing the complex magnetic field as $h=gZ$, where $g$ is the modulus and $\varphi$ the argument.
We numerically calculate the fidelity zeros for a lattice of size $L=10$ at $g=0.5$ and $g=1.5$, with $\varphi$ varying from $0$ to $2\pi$, as shown in Fig.~\ref{Fig:YLES1D}(a) and (b). In both cases, we identify $2L$ fidelity zeros lying on the unit circle, independent of $g$. For $g=0.5$, where the magnitude of the magnetic field is below the critical value $h_{c1}$, the fidelity zeros are uniformly and symmetrically distributed along the circle. In contrast, for $g=1.5$, where the field magnitude exceeds $h_{c1}$, the fidelity zeros are distributed along a circular arc and give rise to branch points, which we designate as fidelity edges.
The above arguments hold for larger lattices, as shown for $L=32$ in Fig.~\ref{Fig:YLES1D}(c) and (d). In this case, the zeros become dense along the circle, while their real part remains bounded by $h_{c1}$.
For $g>h_{c1}$, the fidelity edge in the thermodynamic limit is located at the critical angle $\theta = \arccos(h_{c1}/g)$.

{\bf Two-dimensional quantum Ising model.}  
The quantum Ising chain discussed above provides exact analytical results for both the ground state and fidelity, whereas most strongly correlated systems admit no such solutions.
To further validate our argument, we turn to the two-dimensional (2D) transverse-field quantum Ising model defined by Hamiltonian (\ref{Eq.Ham}) on a square lattice, as depicted in Fig.~\ref{fig:model}(b).
Here, the summation \(\langle i, j \rangle\) runs over all nearest-neighbor pairs along both the $x$ and $y$ directions.
The periodic boundary conditions are given by $\sigma_{L+1,j}^x = \sigma_{1,j}^x$ and $\sigma_{i,L+1}^x = \sigma_{i,1}^x$, where $L$ is the number of sites along each direction of the square lattice, giving a total of $N = L \times L$ sites. The quantum Ising model on a square lattice exhibits a second-order quantum phase transition at the critical point $h_{c2} \simeq 3.044$ \cite{blote2002cluster}.

\begin{figure}[t]
\includegraphics[width=8.8cm]{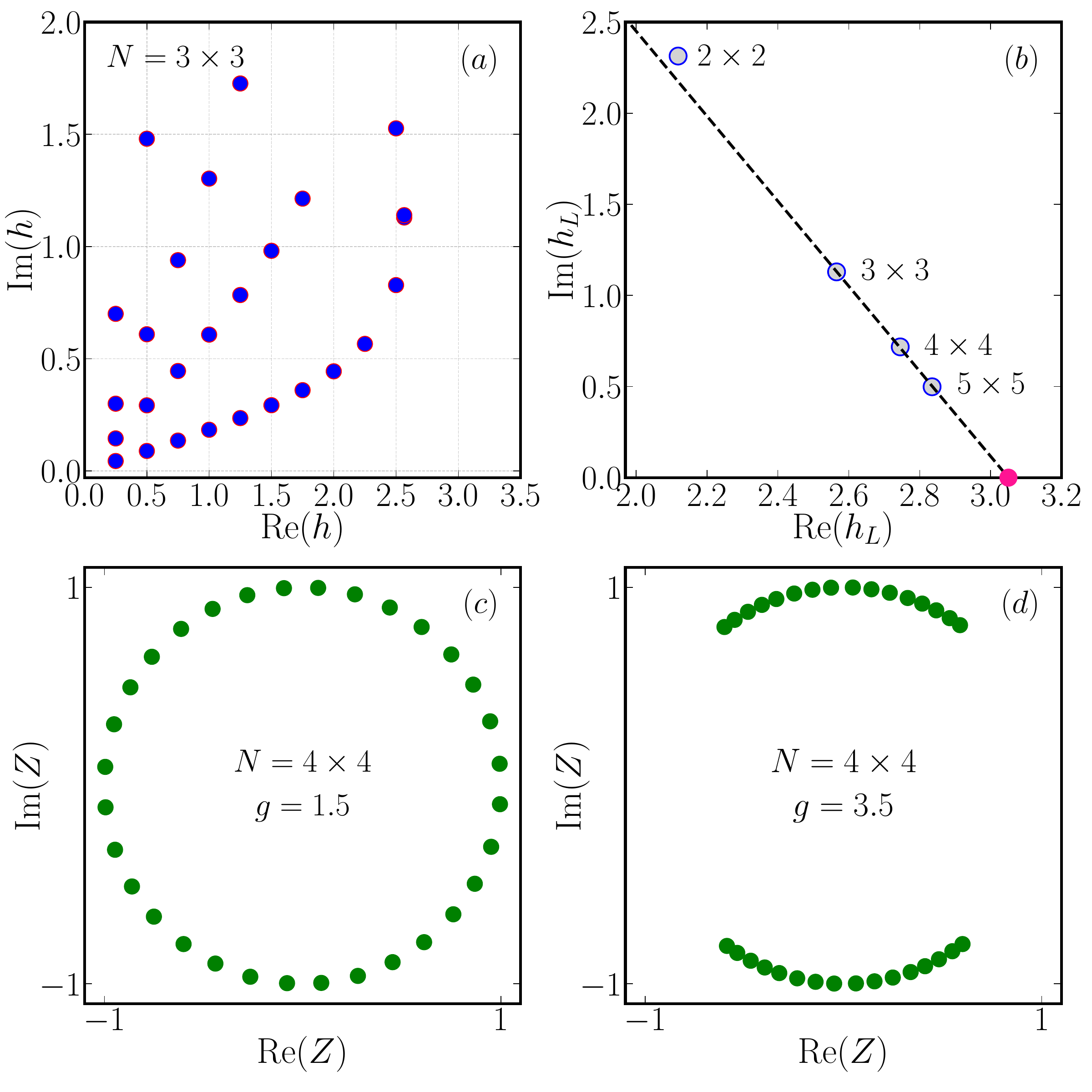} 
\centering
\caption{Fidelity zeros and fidelity edges of 2D quantum Ising model. (a) Distribution of fidelity zeros in the complex plane $h$ with $N=3 \times 3$ and $\delta h = 10^{-4}$. The scatter blue points mark the complex field values at which the fidelity vanishes. The critical point is $\text{Re}(h_{L}) \approx 2.57$ for $N=3 \times 3$. (b) Finite-size scaling of the fidelity-zero positions as a function of system size $L$ ranging from $2$ to $5$. Here, The grey-blue dots denote the complex field values with the largest real and smallest imaginary parts. The black dashed line represents the fitted curve. The pink solid dot on the real axis marks the critical value $h_{c2} = 3.05$, consistent with previous results. (c) and (d) Distributions of fidelity zeros in the complex field plane for $g=1.5$ and $g=3.5$ with $N=4 \times 4$, where the distributions of fidelity zeros are located at the unit circle $Z = e^{i \varphi}$ with $\varphi = [0, 2\pi)$. For $g=3.5$, fidelity edges exist. Here $\delta \varphi = 2\pi/400$ is used for computation of fidelity.}
\label{fig:2D}
\end{figure} 

We numerically compute the ground-state fidelity for the $N = 3 \times 3$ lattice, varying the real component of the complex magnetic field $h$ from $0$ to $3.5$.
The system exhibits fidelity zeros for $\text{Re}(h) < \text{Re}(h_{L})$, while no zeros are observed for $\text{Re}(h) > \text{Re}(h_{L})$, as shown in Fig.~\ref{fig:2D}(a).
This behavior signals a distinct transition at $\text{Re}(h_{L}) \approx 2.57$.
Furthermore, we investigate the finite-size scaling [c.f Fig.~\ref{fig:2D}(b)] of the critical point $h_{L}$ defined as the fidelity zero with the largest real part and smallest imaginary part. 
With increasing lattice size, the real part of $h_{L}$  converges towards a real finite value and the imaginary part vanishes in the thermodynamic limit, giving rise to critical point $h_{c2}=3.05$ with $\nu=0.63$. 
This result is in excellent agreement with established numerical studies based on quantum Monte Carlo \cite{blote2002cluster}, infinite projected entangled-pair states \cite{jordan2008classical,orus2009simulation}, and tensor renormalization group \cite{gu2008tensor,xie2012coarse} methods for 2D transverse-field Ising model, thereby confirming the numerical accuracy of our approach.

Additionally, we investigate the Lee-Yang unit circle theorem by analyzing the distribution of fidelity zeros, where the complex magnetic field is parameterized as $h=gZ$  in terms of the fugacity $Z = e^{i\varphi}$, for a $N=4 \times 4$ lattice system. We observe that all fidelity zeros are located on the unit circle independent of the modulus $g$.
When $g < h_{c2}$, the zeros are uniformly distributed along the unit circle, whereas for $g > h_{c2}$ they condense on two distinct circular arcs, leaving a finite gap between them. This qualitative change in distribution across the critical point demonstrates that the system resides in different quantum phases. 
The critical angle, $\theta = \arccos(h_{c2}/g)$, corresponding to the fidelity edges (the arc boundaries), is obtained from this analysis.
These findings further demonstrate the effectiveness of fidelity zeros as a probe for characterizing quantum phase transitions and establish a direct connection between the fidelity and the partition function.

{\bf Conclusion.} 
In summary, we present a unifying Lee-Yang framework for characterizing quantum phase transitions through fidelity, which play the role of the partition function in classical phase transitions and Loschmidt echo in dynamical quantum phase transitions. We show that fidelity zeros serve as Lee-Yang zeros, comply with the Lee-Yang theorem, and accumulate to form a branch point. Their emergence is explained by transitions between different parity sectors. In addition, we develop a scaling theory for fidelity zeros near the critical point and validate our approach on one- and two-dimensional quantum Ising models.

From the perspective of numerical methods, fidelity zeros and edges can be efficiently and accurately obtained via exact diagonalization, thereby avoiding evaluation of the complex partition function and reducing the numerical instabilities inherent in quantum many-body systems. Furthermore, the results are comparable to those obtained from established numerical methods such as quantum Monte Carlo and tensor-network approaches, indicating that our method has strong potential as a powerful new tool for studying strongly correlated systems.

With the recently developed biorthonormal-block density-matrix renormalization group algorithms \cite{zhong2025density}, fidelity zeros in large-scale non-Hermitian many-body systems can now be computed. It would be interesting to investigate the applicability of this theory to other types of phase transitions, such as Berezinskii-Kosterlitz-Thouless transitions and nonequilibrium quantum phase transitions in the future.

{\bf Acknowledgments.} 
We thank Haiyuan Zou and Wei Chen for valuable discussions.
G.S. is appreciative of support from the NSFC under the Grants NO. 11704186, the Fundamental Research Funds for the Central Universities under the Grants NO. NS2023055. 
This work is partially supported by the High Performance Computing Platform of Nanjing University of Aeronautics and Astronautics.

\bibliography{LYref}

@book{sachdev1999quantum,
  title={Quantum phase transitions},
  author={Sachdev, Subir},
  year={1999},
  publisher={Cambridge University Press}
}

@article{yang1952statistical,
  title={Statistical theory of equations of state and phase transitions. I. Theory of condensation},
  author={Yang, Chen-Ning and Lee, Tsung-Dao},
  journal={Physical Review},
  volume={87},
  number={3},
  pages={404},
  year={1952},
  publisher={APS},
  DOI={https://doi.org/10.1103/PhysRev.87.404}
}

@article{lee1952statistical,
  title={Statistical theory of equations of state and phase transitions. II. Lattice gas and Ising model},
  author={Lee, Tsung-Dao and Yang, Chen-Ning},
  journal={Physical Review},
  volume={87},
  number={3},
  pages={410},
  year={1952},
  publisher={APS},
  DOI={https://doi.org/10.1103/PhysRev.87.410}
}

@book{fisher1965statistical,
  title={The nature of critical points, in Lectures in Theoretical Physics},
  author={Fisher, M. E},
  publisher={vol VIIC, edited by W. E. Brittin, University of Colorado Press},
  year={1965}
}

@article{heyl2013dynamical,
  title={Dynamical quantum phase transitions in the transverse-field Ising model},
  author={Heyl, Markus and Polkovnikov, Anatoli and Kehrein, Stefan},
  journal={Physical Review Letters},
  volume={110},
  number={13},
  pages={135704},
  year={2013},
  publisher={APS},
  DOI={https://doi.org/10.1103/PhysRevLett.110.135704}
}

@article{heyl2018dynamical,
  title={Dynamical quantum phase transitions: a review},
  author={Heyl, Markus},
  journal={Reports on Progress in Physics},
  volume={81},
  number={5},
  pages={054001},
  year={2018},
  publisher={IOP Publishing},
  DOI={10.1088/1361-6633/aaaf9a}
}

@article{fisher1978yang,
  title={Yang-Lee edge singularity and $\phi$ 3 field theory},
  author={Fisher, Michael E},
  journal={Physical Review Letters},
  volume={40},
  number={25},
  pages={1610},
  year={1978},
  publisher={APS},
  DOI={https://doi.org/10.1103/PhysRevLett.40.1610}
}

@article{kortman1971density,
  title={Density of zeros on the Lee-Yang circle for two Ising ferromagnets},
  author={Kortman, Peter J and Griffiths, Robert B},
  journal={Physical Review Letters},
  volume={27},
  number={21},
  pages={1439},
  year={1971},
  publisher={APS},
  DOI={https://doi.org/10.1103/PhysRevLett.27.1439}
}

@article{cardy1985conformal,
  title={Conformal invariance and the Yang-Lee edge singularity in two dimensions},
  author={Cardy, John L},
  journal={Physical Review Letters},
  volume={54},
  number={13},
  pages={1354},
  year={1985},
  publisher={APS},
  DOI={https://doi.org/10.1103/PhysRevLett.54.1354}
}

@article{ananikian2015imaginary,
  title={Imaginary magnetic fields in the real world},
  author={Ananikian, Nerses and Kenna, Ralph},
  journal={Physics},
  volume={8},
  pages={2},
  year={2015},
  publisher={APS},
  DOI={http://link.aps.org/doi/10.1103/Physics.8.2}
}

@article{peng2015experimental,
  title={Experimental observation of Lee-Yang zeros},
  author={Peng, Xinhua and Zhou, Hui and Wei, Bo-Bo and Cui, Jiangyu and Du, Jiangfeng and Liu, Ren-Bao},
  journal={Physical Review Letters},
  volume={114},
  number={1},
  pages={010601},
  year={2015},
  publisher={APS},
  DOI={https://doi.org/10.1103/PhysRevLett.114.010601}
}

@article{binek1998density,
  title={Density of zeros on the Lee-Yang circle obtained from magnetization data of a two-dimensional Ising ferromagnet},
  author={Binek, Ch},
  journal={Physical Review Letters},
  volume={81},
  number={25},
  pages={5644},
  year={1998},
  publisher={APS},
  DOI={https://doi.org/10.1103/PhysRevLett.81.5644}
}

@article{binek2001yang,
  title={Yang-Lee edge singularities determined from experimentalhigh-field magnetization data},
  author={Binek, Ch and Kleemann, Wolfgang and Katori, H Aruga},
  journal={Journal of Physics: Condensed Matter},
  volume={13},
  number={35},
  pages={L811},
  year={2001},
  publisher={IOP Publishing},
  DOI={10.1088/0953-8984/13/35/103}
}

@article{wei2012lee,
  title={Lee-Yang zeros and critical times in decoherence of a probe spin coupled to a bath},
  author={Wei, Bo-Bo and Liu, Ren-Bao},
  journal={Physical Review Letters},
  volume={109},
  number={18},
  pages={185701},
  year={2012},
  publisher={APS},
  DOI={https://doi.org/10.1103/PhysRevLett.109.185701}
}

@article{bergholtz2021exceptional,
  title={Exceptional topology of non-Hermitian systems},
  author={Bergholtz, Emil J and Budich, Jan Carl and Kunst, Flore K},
  journal={Reviews of Modern Physics},
  volume={93},
  number={1},
  pages={015005},
  year={2021},
  publisher={APS},
  DOI={https://doi.org/10.1103/RevModPhys.93.015005}
}

@article{ashida2020non,
  title={Non-hermitian physics},
  author={Ashida, Yuto and Gong, Zongping and Ueda, Masahito},
  journal={Advances in Physics},
  volume={69},
  number={3},
  pages={249--435},
  year={2020},
  publisher={Taylor \& Francis},
  DOI={https://doi.org/10.1080/00018732.2021.1876991}
}

@article{yao2018edge,
  title={Edge states and topological invariants of non-Hermitian systems},
  author={Yao, Shunyu and Wang, Zhong},
  journal={Physical Review Letters},
  volume={121},
  number={8},
  pages={086803},
  year={2018},
  publisher={APS},
  DOI={https://doi.org/10.1103/PhysRevLett.121.086803}
}

@article{yokomizo2019non,
  title={Non-Bloch band theory of non-Hermitian systems},
  author={Yokomizo, Kazuki and Murakami, Shuichi},
  journal={Physical Review Letters},
  volume={123},
  number={6},
  pages={066404},
  year={2019},
  publisher={APS},
  DOI={https://doi.org/10.1103/PhysRevLett.123.066404}
}

@article{yang2020non,
  title={Non-Hermitian bulk-boundary correspondence and auxiliary generalized Brillouin zone theory},
  author={Yang, Zhesen and Zhang, Kai and Fang, Chen and Hu, Jiangping},
  journal={Physical Review Letters},
  volume={125},
  number={22},
  pages={226402},
  year={2020},
  publisher={APS},
  DOI={https://doi.org/10.1103/PhysRevLett.125.226402}
}

@article{okuma2020topological,
  title={Topological origin of non-Hermitian skin effects},
  author={Okuma, Nobuyuki and Kawabata, Kohei and Shiozaki, Ken and Sato, Masatoshi},
  journal={Physical Review Letters},
  volume={124},
  number={8},
  pages={086801},
  year={2020},
  publisher={APS},
  DOI={https://doi.org/10.1103/PhysRevLett.124.086801}
}

@article{zhang2020correspondence,
  title={Correspondence between winding numbers and skin modes in non-hermitian systems},
  author={Zhang, Kai and Yang, Zhesen and Fang, Chen},
  journal={Physical Review Letters},
  volume={125},
  number={12},
  pages={126402},
  year={2020},
  publisher={APS},
  DOI={https://doi.org/10.1103/PhysRevLett.125.126402}
}

@article{lin2023topological,
  title={Topological non-Hermitian skin effect},
  author={Lin, Rijia and Tai, Tommy and Li, Linhu and Lee, Ching Hua},
  journal={Frontiers of Physics},
  volume={18},
  number={5},
  pages={53605},
  year={2023},
  publisher={Springer},
  DOI={https://doi.org/10.1007/s11467-023-1309-z}
}

@article{gao2024experimental,
  title={Experimental Observation of the Yang-Lee Quantum Criticality in Open Quantum Systems},
  author={Gao, Huixia and Wang, Kunkun and Xiao, Lei and Nakagawa, Masaya and Matsumoto, Norifumi and Qu, Dengke and Lin, Haiqing and Ueda, Masahito and Xue, Peng},
  journal={Physical Review Letters},
  volume={132},
  number={17},
  pages={176601},
  year={2024},
  publisher={APS},
  DOI={https://doi.org/10.1103/PhysRevLett.132.176601}
}

@article{lan2024experimental,
  title={Experimental Investigation of Lee--Yang Criticality Using Non-Hermitian Quantum System},
  author={Lan, Ziheng and Liu, Wenquan and Wu, Yang and Ye, Xiangyu and Yang, Zhesen and Duan, Chang-Kui and Wang, Ya and Rong, Xing},
  journal={Chinese Physics Letters},
  volume={41},
  number={5},
  pages={050301},
  year={2024},
  publisher={IOP Publishing},
  DOI={10.1088/0256-307X/41/5/050301}
}

@article{brandner2017experimental,
  title={Experimental determination of dynamical Lee-Yang zeros},
  author={Brandner, Kay and Maisi, Ville F and Pekola, Jukka P and Garrahan, Juan P and Flindt, Christian},
  journal={Physical Review Letters},
  volume={118},
  number={18},
  pages={180601},
  year={2017},
  publisher={APS},
  DOI={https://doi.org/10.1103/PhysRevLett.118.180601}
}

@article{jurcevic2017direct,
  title={Direct observation of dynamical quantum phase transitions in an interacting many-body system},
  author={Jurcevic, P and Shen, H and Hauke, P and Maier, C and Brydges, T and Hempel, C and Lanyon, B. P. and Heyl, Markus and Blatt, R and Roos, C. F.},
  journal={Physical Review Letters},
  volume={119},
  number={8},
  pages={080501},
  year={2017},
  publisher={APS},
  DOI={https://doi.org/10.1103/PhysRevLett.119.080501}
}

@article{flaschner2018observation,
  title={Observation of dynamical vortices after quenches in a system with topology},
  author={Fl{\"a}schner, N and Vogel, D and Tarnowski, M and Rem, BS and L{\"u}hmann, D-S and Heyl, M and Budich, JC and Mathey, L and Sengstock, K and Weitenberg, C},
  journal={Nature Physics},
  volume={14},
  number={3},
  pages={265},
  year={2018},
  publisher={Nature Publishing Group},
  DOI={https://doi.org/10.1038/s41567-017-0013-8}
}

@article{xu2020measuring,
  title={Measuring a dynamical topological order parameter in quantum walks},
  author={Xu, Xiao-Ye and Wang, Qin-Qin and Heyl, Markus and Budich, Jan Carl and Pan, Wei-Wei and Chen, Zhe and Jan, Munsif and Sun, Kai and Xu, Jin-Shi and Han, Yong-Jian and  Li, Chuan-Feng and Guo, Guang-Can},
  journal={Light: Science \& Applications},
  volume={9},
  number={1},
  pages={1-11},
  year={2020},
  publisher={Nature Publishing Group},
  DOI={https://doi.org/10.1038/s41377-019-0237-8}
}

@article{wang2019simulating,
  title={Simulating dynamic quantum phase transitions in photonic quantum walks},
  author={Wang, Kunkun and Qiu, Xingze and Xiao, Lei and Zhan, Xiang and Bian, Zhihao and Yi, Wei and Xue, Peng},
  journal={Physical Review Letters},
  volume={122},
  number={2},
  pages={020501},
  year={2019},
  publisher={APS},
  DOI={https://doi.org/10.1103/PhysRevLett.122.020501}
}

@article{tian2019observation,
  title={Observation of dynamical phase transitions in a topological nanomechanical system},
  author={Tian, Tian and Ke, Yongguan and Zhang, Liang and Lin, Shaochun and Shi, Zhifu and Huang, Pu and Lee, Chaohong and Du, Jiangfeng},
  journal={Physical Review B},
  volume={100},
  number={2},
  pages={024310},
  year={2019},
  publisher={APS},
  DOI={https://doi.org/10.1103/PhysRevB.100.024310}
}

@article{guo2019observation,
  title={Observation of a Dynamical Quantum Phase Transition by a Superconducting Qubit Simulation},
  author={Guo, Xue-Yi and Yang, Chao and Zeng, Yu and Peng, Yi and Li, He-Kang and Deng, Hui and Jin, Yi-Rong and Chen, Shu and Zheng, Dongning and Fan, Heng},
  journal={Physical Review Applied},
  volume={11},
  number={4},
  pages={044080},
  year={2019},
  publisher={APS},
  DOI={https://doi.org/10.1103/PhysRevApplied.11.044080}
}

@article{nie2020experimental,
  title={Experimental observation of equilibrium and dynamical quantum phase transitions via out-of-time-ordered correlators},
  author={Nie, Xinfang and Wei, Bo-Bo and Chen, Xi and Zhang, Ze and Zhao, Xiuzhu and Qiu, Chudan and Tian, Yu and Ji, Yunlan and Xin, Tao and Lu, Dawei and Li, Jun},
  journal={Physical Review Letters},
  volume={124},
  number={25},
  pages={250601},
  year={2020},
  publisher={APS},
  DOI={https://doi.org/10.1103/PhysRevLett.124.250601}
}

@article{tian2020observation,
  title={Observation of dynamical quantum phase transitions with correspondence in an excited state phase diagram},
  author={Tian, T and Yang, H-X and Qiu, L-Y and Liang, H-Y and Yang, Y-B and Xu, Y and Duan, L-M},
  journal={Physical Review Letters},
  volume={124},
  number={4},
  pages={043001},
  year={2020},
  publisher={APS},
  DOI={https://doi.org/10.1103/PhysRevLett.124.043001}
}

@article{wu2024indication,
  title={Indication of critical scaling in time during the relaxation of an open quantum system},
  author={Wu, Ling-Na and Nettersheim, Jens and Fe{\ss}, Julian and Schnell, Alexander and Burgardt, Sabrina and Hiebel, Silvia and Adam, Daniel and Eckardt, Andr{\'e} and Widera, Artur},
  journal={Nature Communications},
  volume={15},
  number={1},
  pages={1714},
  year={2024},
  publisher={Nature Publishing Group UK London},
  DOI={https://doi.org/10.1038/s41467-024-46054-9}
}

@article{kist2021lee,
  title={Lee-Yang theory of criticality in interacting quantum many-body systems},
  author={Kist, Timo and Lado, Jose L and Flindt, Christian},
  journal={Physical Review Research},
  volume={3},
  number={3},
  pages={033206},
  year={2021},
  publisher={APS},
  DOI={https://doi.org/10.1103/PhysRevResearch.3.033206}
}

@article{vecsei2022lee,
  title={Lee-Yang theory of the two-dimensional quantum Ising model},
  author={Vecsei, Pascal M and Lado, Jose L and Flindt, Christian},
  journal={Physical Review B},
  volume={106},
  number={5},
  pages={054402},
  year={2022},
  publisher={APS},
  DOI={https://doi.org/10.1103/PhysRevB.106.054402}
}

@article{vecsei2023lee,
  title={Lee-Yang theory of quantum phase transitions with neural network quantum states},
  author={Vecsei, Pascal M and Flindt, Christian and Lado, Jose L},
  journal={Physical Review Research},
  volume={5},
  number={3},
  pages={033116},
  year={2023},
  publisher={APS},
  DOI={https://doi.org/10.1103/PhysRevResearch.5.033116}
}

@article{liu2024imaginary,
  title={Imaginary-temperature zeros for quantum phase transitions},
  author={Liu, Jinghu and Yin, Shuai and Chen, Li},
  journal={Physical Review B},
  volume={110},
  number={13},
  pages={134313},
  year={2024},
  publisher={APS},
  DOI={https://doi.org/10.1103/PhysRevB.110.134313}
}

@article{liu2024exact,
  title={Exact Fisher zeros and thermofield dynamics across a quantum critical point},
  author={Liu, Yang and Lv, Songtai and Meng, Yuchen and Tan, Zefan and Zhao, Erhai and Zou, Haiyuan},
  journal={Physical Review Research},
  volume={6},
  number={4},
  pages={043139},
  year={2024},
  publisher={APS},
  DOI={https://doi.org/10.1103/PhysRevResearch.6.043139}
}

@article{vecsei2025lee,
  title={Lee-Yang formalism for phase transitions of interacting fermions using tensor networks},
  author={Vecsei, Pascal M and Lado, Jose L and Flindt, Christian},
  journal={Physical Review B},
  volume={111},
  number={7},
  pages={075134},
  year={2025},
  publisher={APS},
  DOI={https://doi.org/10.1103/PhysRevB.111.075134}
}

@article{li2023yang,
  title={Yang-Lee zeros, semicircle theorem, and nonunitary criticality in Bardeen-Cooper-Schrieffer superconductivity},
  author={Li, Hongchao and Yu, Xie-Hang and Nakagawa, Masaya and Ueda, Masahito},
  journal={Physical Review Letters},
  volume={131},
  number={21},
  pages={216001},
  year={2023},
  publisher={APS},
  DOI={https://doi.org/10.1103/PhysRevLett.131.216001}
}

@article{li2025yang,
  title={Yang-Lee zeros in quantum phase transitions: An entanglement perspective},
  author={Li, Hongchao},
  journal={Physical Review B},
  volume={111},
  number={4},
  pages={045139},
  year={2025},
  publisher={APS},
  DOI={https://doi.org/10.1103/PhysRevB.111.045139}
}

@article{meng2025detecting,
  title={Detecting Many-Body Scars from Fisher Zeros},
  author={Meng, Yuchen and Lv, Songtai and Liu, Yang and Tan, Zefan and Zhao, Erhai and Zou, Haiyuan},
  journal={Physical Review Letters},
  volume={135},
  number={7},
  pages={070402},
  year={2025},
  publisher={APS},
  DOI={https://doi.org/10.1103/glc5-hv2m}
}

@article{wang2024quantum,
  title={Quantum Supercritical Crossover with Dynamical Singularity},
  author={Wang, Junsen and Lv, Enze and Li, Xinyang and Jin, Yuliang and Li, Wei},
  journal={arXiv:2407.05455},
  year={2024},
  url={https://arxiv.org/abs/2407.05455}
}

@article{he2025revisiting,
  title={Revisiting the Fermion Sign Problem from the Structure of Lee-Yang Zeros. I. The Form of Partition Function for Indistinguishable Particles and Its Zeros at 0 K},
  author={He, Ran-Chen and Zeng, Jia-Xi and Yang, Shu and Wang, Cong and Ye, Qi-Jun and Li, Xin-Zheng},
  journal={arXiv:2507.22779},
  year={2025},
  url={https://arxiv.org/abs/2507.22779}
}

@article{mbeng2024quantum,
  title={The quantum Ising chain for beginners},
  author={Mbeng, Glen Bigan and Russomanno, Angelo and Santoro, Giuseppe E},
  journal={SciPost Physics Lecture Notes},
  pages={082},
  year={2024},
  DOI={10.21468/SciPostPhysLectNotes.82}
}

@article{sun2023aufbau,
  title={Aufbau principle for non-hermitian systems},
  author={Sun, Gaoyong and Kou, Su-Peng},
  journal={arXiv:2307.04696},
  year={2023},
  url={https://arxiv.org/abs/2307.04696}
}

@article{gu2010fidelity,
  title={Fidelity approach to quantum phase transitions},
  author={Gu, Shi-Jian},
  journal={International Journal of Modern Physics B},
  volume={24},
  number={23},
  pages={4371--4458},
  year={2010},
  publisher={World Scientific},
  DOI={https://doi.org/10.1142/S0217979210056335}
}

@article{you2007fidelity,
  title={Fidelity, dynamic structure factor, and susceptibility in critical phenomena},
  author={You, Wen-Long and Li, Ying-Wai and Gu, Shi-Jian},
  journal={Physical Review E},
  volume={76},
  number={2},
  pages={022101},
  year={2007},
  publisher={APS},
  DOI={https://doi.org/10.1103/PhysRevE.76.022101}
}

@article{zeng2024exact,
  title={Exact zeros of fidelity in finite-size systems as a signature for probing quantum phase transitions},
  author={Zeng, Yumeng and Zhou, Bozhen and Chen, Shu},
  journal={Physical Review E},
  volume={109},
  number={6},
  pages={064130},
  year={2024},
  publisher={APS},
  DOI={https://doi.org/10.1103/PhysRevE.109.064130}
}

@article{blote2002cluster,
  title={Cluster Monte Carlo simulation of the transverse Ising model},
  author={Bl{\"o}te, Henk WJ and Deng, Youjin},
  journal={Physical Review E},
  volume={66},
  number={6},
  pages={066110},
  year={2002},
  publisher={APS},
  DOI={https://doi.org/10.1103/PhysRevE.66.066110}
}

@article{jordan2008classical,
  title={Classical simulation of infinite-size quantum lattice systems in two spatial dimensions},
  author={Jordan, Jacob and Or{\'u}s, Roman and Vidal, Guifre and Verstraete, Frank and Cirac, J Ignacio},
  journal={Physical Review Letters},
  volume={101},
  number={25},
  pages={250602},
  year={2008},
  publisher={APS},
  DOI={10.1103/PhysRevLett.101.250602}
}

@article{orus2009simulation,
  title={Simulation of two-dimensional quantum systems on an infinite lattice revisited: Corner transfer matrix for tensor contraction},
  author={Or{\'u}s, Rom{\'a}n and Vidal, Guifr{\'e}},
  journal={Physical Review B},
  volume={80},
  number={9},
  pages={094403},
  year={2009},
  publisher={APS},
  DOI={https://doi.org/10.1103/PhysRevB.80.094403}
}

@article{gu2008tensor,
  title={Tensor-entanglement renormalization group approach as a unified method for symmetry breaking and topological phase transitions},
  author={Gu, Zheng-Cheng and Levin, Michael and Wen, Xiao-Gang},
  journal={Physical Review B},
  volume={78},
  number={20},
  pages={205116},
  year={2008},
  publisher={APS},
  DOI={https://doi.org/10.1103/PhysRevB.78.205116}
}

@article{xie2012coarse,
  title={Coarse-graining renormalization by higher-order singular value decomposition},
  author={Xie, Zhi-Yuan and Chen, Jing and Qin, Ming-Pu and Zhu, Jing W and Yang, Li-Ping and Xiang, Tao},
  journal={Physical Review B},
  volume={86},
  number={4},
  pages={045139},
  year={2012},
  publisher={APS},
  DOI={https://doi.org/10.1103/PhysRevB.86.045139}
}

@article{zhong2025density,
  title={Density matrix renormalization group algorithm for non-Hermitian systems},
  author={Zhong, Peigeng and Pan, Wei and Lin, Haiqing and Wang, Xiaoqun and Hu, Shijie},
  journal={Physical Review Letters},
  volume={135},
  number={10},
  pages={106502},
  year={2025},
  publisher={APS},
  DOI={https://doi.org/10.1103/5vnl-w9p4}
}


\end{document}